\documentclass[10pt,letterpaper]{article}
\usepackage[top=0.85in,left=2.75in,footskip=0.75in]{geometry}

\usepackage{changepage}

\usepackage[utf8]{inputenc}

\usepackage{textcomp,marvosym}

\usepackage{fixltx2e}

\usepackage{amsmath,amssymb}

\usepackage{cite}

\usepackage{nameref,hyperref}

\usepackage[right]{lineno}

\usepackage{microtype}
\DisableLigatures[f]{encoding = *, family = * }

\usepackage{rotating}

\usepackage{graphicx}
\usepackage{epstopdf}
\raggedright
\usepackage[aboveskip=1pt,labelfont=bf,labelsep=period,justification=raggedright,singlelinecheck=off]{caption}

\usepackage{ragged2e}

\makeatletter
\renewcommand{\@biblabel}[1]{\quad#1.}
\makeatother

\date{}

\usepackage{lastpage,fancyhdr,graphicx}
\fancyheadoffset[L]{2.25in}
\fancyfootoffset[L]{2.25in}
 \usepackage{amsfonts}
 \usepackage[section]{algorithm}
\floatname{algorithm}{Algorithme}
\usepackage{algorithmicx}
\usepackage{color,colortbl,graphics}

\begin{document}
\vspace*{0.35in}

\begin{flushleft}
{\Large

\textbf\newline{Modeling the Influence of Local Environmental Factors
on Malaria Transmission in Benin and Its Implications for
Cohort Study}
}
\newline

 \textbf{Gilles Cottrell $^{1,2,3,4,5 *,@}$ , Bienvenue Kouwaye $^{1,2,3,4,5,6,@}$ , Charlotte Pierrat $^{1,2,3,4,5,6,7,8}$ , Agn\`es le Port $^{2,3}$ , Aziz
Boura\"ima $^{4,5}$ , No\"el Fonton $^{6}$ , Mahouton Norbert Hounkonnou $^{6}$ , Achille Massougbodji $^{5,9}$ , Vincent Corbel $^{7}$ ,
Andr\'e Garcia $^{2,3}$}

\textbf{1} Institut de Recherche pour le D\'eveloppement (IRD), M\`ere et Enfant Face aux Infections Tropicales, Cotonou, Benin,
\textbf{2} M\`ere et Enfant Face aux Infections Tropicales, Paris, France,
\textbf{3} Facult\'e de Pharmacie, Universit\'e Paris Descartes, Paris, France, 
\textbf{4} Institut des Sciences Biom\'edicales
Applique\'es (ISBA), Cotonou, Benin, 
\textbf{5} Centre d’Etudes et de Recherche sur le Paludisme Associ\'e \`a la Grossesse et \`a l’Enfant (CERPAGE), Contonou, Benin,
\textbf{6} Universit\'e
d’Abomey-Calavi, International Chair in Mathematical Physics and Applications (ICMPA-UNESCO Chair), Cotonou, Benin,
\textbf{7} Institut de Recherche pour le D\'eveloppement
(IRD), Maladies Infectieuses et Vecteurs, Ecologie, G\'en\'etique, Evolution et Contr\^ole (MIVEGEC, UM1-CNRS 5290-IRD 224), Centre de Recherche Entomologique de Cotonou
(CREC), Cotonou, B\'enin, 
\textbf{8} Universit\'e Paris I Panth\'eon-Sorbonne, Paris, France, 
\textbf{9} Facult\'e des Sciences de la Sant\'e (FSS), Cotonou, Benin\\

* E-mail: Gilles.Cottrell@ird.fr\\
$@$ These authors contributed equally to this work.\\

%
%






\end{flushleft}

\justifying
\section*{Abstract} 
Malaria remains endemic in tropical areas, especially in Africa. For the evaluation of new tools and to further our
understanding of host-parasite interactions, knowing the environmental risk of transmission—even at a very local scale—is
essential. The aim of this study was to assess how malaria transmission is influenced and can be predicted by local climatic
and environmental factors. As the entomological part of a cohort study of 650 newborn babies in nine villages in the Tori
Bossito district of Southern Benin between June 2007 and February 2010, human landing catches were performed to assess
the density of malaria vectors and transmission intensity. Climatic factors as well as household characteristics were recorded
throughout the study. Statistical correlations between Anopheles density and environmental and climatic factors were
tested using a three-level Poisson mixed regression model. The results showed both temporal variations in vector density
(related to season and rainfall), and spatial variations at the level of both village and house. These spatial variations could be
largely explained by factors associated with the house’s immediate surroundings, namely soil type, vegetation index and
the proximity of a watercourse. Based on these results, a predictive regression model was developed using a leave-one-out
method, to predict the spatiotemporal variability of malaria transmission in the nine villages. This study points up the
importance of local environmental factors in malaria transmission and describes a model to predict the transmission risk of
individual children, based on environmental and behavioral characteristics.\\
\newpage
\noindent \textbf{Citation} : Cottrell G, Kouwaye B, Pierrat C, le Port A, Boura\"ima A, et al. (2012) Modeling the Influence of Local Environmental Factors on Malaria Transmission in
Benin and Its Implications for Cohort Study. PLoS ONE 7(1): e28812. doi:10.1371/journal.pone.0028812\\
\noindent \textbf{Editor} : Clive Shiff, Johns Hopkins University, United States of America\\

\noindent \textbf{Received} July 27, 2011; \textbf{Accepted} November 15, 2011; \textbf{Published} January 4, 2012\\
\noindent \textbf{Copyright} : $\copyright$ 2012 Cottrell et al. This is an open-access article distributed under the terms of the Creative Commons Attribution License, which permits
unrestricted use, distribution, and reproduction in any medium, provided the original author and source are credited.\\

\noindent \textbf{Funding}: Agence Nationale de la Recherche (ANR www.agence-nationale-recherche.fr) project SEST 2006 (040-01); Minist\`ere des Affaires Etrang\`eres (France,
www.diplomatie.gouv.fr) project REFS Nu2006-22; and the Institut de Recherche pour le Developpement, which financially and materially supported the work. The
funders had no role in study design, data collection and analysis, decision to publish, or preparation of the manuscript.\\
\noindent \textbf{Competing Interests}: The authors have declared that no competing interests exist.\\
\noindent * E-mail: Gilles.Cottrell@ird.fr\\
\noindent These authors contributed equally to this work.

\section*{Introduction}
Malaria remains endemic in sub-Saharan Africa although dramatic
declines in morbidity have been reported in the last five years across a
range of settings [1,2,3]. This is associated with the distribution of
long-lasting insecticide-treated mosquito nets and a switch to first line
artemisinin-based combination therapy (ACT). Nonetheless, the
disease’s burden is still high in Africa where it is a leading cause of
mortality, especially in children of under five years of age [4]: new
tools—a vaccine, effective drugs, better insecticides—are still needed
together with strategies for their use and evaluation. In addition,
improving our understanding of host-parasite interactions is a priority.
Accurately assessing the local risk of transmission is fundamental
to the development of a malaria control program. In Africa,
transmission levels vary enormously and transmission may be either seasonal or perennial [5]. Differences exist not just between
different regions but also at the very local level [6,7,8]. Key
determinants of local transmission intensity [9,10] include vector
profile, ecology and seasonality [11,12], all of which will affect the
efficacy of control operations. The results of recent studies in two
different countries (Ghana and Gabon) point up the importance of
high-resolution analysis of local variations when designing and
monitoring a malaria control operation [12,13].
Recent findings showed that small-scale differences within an
area may have important consequences when it comes to studying
individual responses to a risk of infection or to an intervention, e.g.
responses to vaccination may be quite different in children who
have not been exposed to the antigen to the same extent [14,15].
Therefore, localized variations ought to be taken into account
when considering the risk of infection in a population and the determinants of individual variability (i.e. behavior, physiology
and genetics).
This applies to the consequences of placenta-associated malaria
(PAM) on the development of specific immunity to P. falciparum
and the lag before appearance of the first infection in newborns.
Four studies showed that children born to mothers with PAM have
a higher risk of infection during their first months of life, pointing
to the phenomenon of immune tolerance [16,17,18,19]. However,
these studies failed to take spatiotemporal variations into account
with no entomological or environmental data input into the
analyses [20,21]. Thus, it cannot be ruled out that differences in
transmission risk may have affected outcomes, i.e. no conclusion
can be drawn about immune tolerance from cohort studies unless
information about spatiotemporal variations in malaria transmis-
sion is included in the analysis.
This article describes- through the, statistical analysis of the
entomological and environmental data of a cohort study
conducted in Southern Benin- a new approach to predict the risk
of malaria transmission in cohort studies.

\section*{Methods}
\subsection*{Ethics}
A written informed consent was obtained from all participants
involved in the study. The study protocol was approved by the
Ethics Committee of the University of Abomey-Calavi (Facult\'e des
Sciences de la Sant\'e; FSS) in Benin and the Consultative Committee
of Ethics of Institute of Development Research (IRD).
\subsection*{Study area}
The study was conducted in the district of Tori-Bossito
(Republic of Benin), between July 2007 and July 2009. Tori
Bossito is on the coastal plain of Southern Benin, 40 kilometers
north-east of Cotonou. This area has a subtropical climate and
during the study the rainy season lasted from May to October.
Average monthly temperatures varied between 27uC and 31uC.
The original equatorial forest has been cleared and the vegetation
is characterized by bushes with sparse trees, a few oil palm
plantations and farms. The study area contained nine villages
(Avam\`ecentre, Gb\'edjougo, Houngo, Anav\'e, Dohinoko, Gb\'etaga,
Tori Cada Centre, Z\`eb\`e and Zoungoudo). Tori Bossito was
recently classified as mesoendemic with a clinical malaria
incidence of about 1.5 episodes per child per year [22].
Pyrethroid-resistant malaria vectors are present [8].
Mosquito collection and identification
Entomological surveys based on human landing catches (HLC)
were performed in the nine villages every six weeks for two years
(July 2007 to July 2009). Mosquitoes were collected at four catch
houses in each village over three successive nights (four indoors and
four outdoors, i.e. a total of 216 nights every six weeks in the nine
villages). Five catch sites had to be changed in the course of the study
(2 in Gbedjougo, 1 in Avame`, 1 in Cada, 1 in Dohinoko) and a total
of 19 data collections were performed in the field between July 2007
and July 2009. In total, data from 41 catch sites are available.
Each collector caught all mosquitoes landing on the lower legs
and feet between 10 pm and 6 am. All mosquitoes were held in
bags labeled with the time of collection. The following morning,
mosquitoes were identified on the basis of morphological criteria
[23,24]. All An. gambiae complex and An. funestus mosquitoes were
stored in individual tubes with silica gel and preserved at 220uC.
P. falciparum infection rates were then determined on the head and
thorax of individual anopheline specimens by CSP-ELISA [25].
Environmental and behavioral data
Rainfall was recorded twice a day with a pluviometer in each
village. In and around each catch site, the following information
was systematically collected: (1) type of soil (dry lateritic or humid
hydromorphic)—assessed using a soil map of the area (map IGN
B\'enin at 1/200 000 e , sheets NB-31-XIV and NB-31-XV, 1968)
that was georeferenced and input into a GIS; (2) presence of areas
where building constructions are ongoing with tools or holes
representing potential breeding habitats for anopheles; (3)
presence of abandoned objects (or ustensils) susceptible to be used
as oviposition sites for female mosquitoes; (4) a watercourse
nearby; (5) number of windows and doors; (6) type of roof (straw or
metal); (7) number of inhabitants; (8) ownership of a bed-net or (9)
insect repellent; and (10) normalized difference vegetation index
(NDVI) which was estimated for 100 meters around the catch site
with a SPOT 5 High Resolution (10 m colors) satellite image
(Image Spot5, CNES, 2003, distribution SpotImage S.A) with
assessment of the chlorophyll density of each pixel of the image.
Due to logistical problems, rainfall measurements are only
available after the second entomological survey. Consequently, we
excluded the first and second surveys (performed in July and
August 2007 respectively) from the statistical analyses. However,
the results of all 19 entomological catches were included in the
descriptive part of the results.

\subsection*{Statistical analysis}
The statistical analysis was conducted in two phases.
First an explanatory regression model was constructed to determine
the correlation between Anopheles density and the above-mentioned
environmental factors. On the basis of these results, a predictive model
was constructed to predict spatiotemporal malaria transmission in
houses for which environmental but not entomological data were
available. The error distribution of this model was compared with that
of a simple ‘‘pragmatic’’ model based on real entomological data.\\

\textbf{ Variables}. The dependent variable was the number of
Anopheles collected in a house over the three nights of each
catch, and the explanatory variables were the environmental
factors, i.e. the mean rainfall between two catches (classified
according to quartile), the number of rainy days in the ten days
before the catch (3 classes [0–1], [2–4], .4 days), the season
during which the catch was carried out (4 classes: end of the dry
season—February to April; beginning of the rainy season—May to
July; end of the rainy season—August to October; beginning of the
dry season—November to January), the type of soil 100 meters
around the house (dry or humid), the presence of constructions
within 100 meters of the house (yes/no), the presence of
abandoned tools within 100 meters of the house (yes/no), the
presence of a watercourse within 500 meters of the house (yes/no),
NDVI 100 meters around the house (classified according to
quartile), the type of roof (straw or sheet metal), the number of
windows (classified according to quartile), the ownership of bed nets (yes/no), the use of insect repellent (yes/no) and the number
of inhabitants in the house (classified according to quartile).\\
\textbf{ Explanatory model}. Since the dependent variable was a
count, in order to take into account the hierarchical structure of the
data (repeated catches in the same house, four sites per village) with
correlation possible between the entomological measurements, a
Poisson mixed model was constructed with three random intercepts
at the village, site and catch levels, i.e. for the k th catch in the j th site
in the i th village:
\begin{equation}
 \label{model_glmm_explicatif}
 \ln [\mathbb{E}(Y_{ijk}|\, a_i, b_{ij}, c_{ijk}; \beta)] = \beta_{0}+\sum_{l=1}^{p}\beta_{l}X_{ijkl}+a_i+b_{ij}+c_{ijk}
\end{equation}
where Y is the number of collected Anopheles, X is a p-vector of
environmental variables, b is a (p+1)-vector of the model’s
parameters (including the fixed intercept b 0 ), a i is the random
intercept at the village level, b ij is the random intercept at the site
level and c ijk is the random intercept at the catch level. It can be
shown that in this model 
\begin{equation}
 \label{model_glmm_variance_experance}
 Var(Y_{ijk}|\, a_i, b_{ij})=\mathbb{E}(Y_{ijk}|\, a_i, b_{ij})+(\mathbb{E}(Y_{ijk}|\, a_i, b_{ij}))^2[exp(\sigma_c^2) - 1] 
\end{equation}

Then $Var(Y_{ijk}|\, a_i, b_{ij}) \geq \mathbb{E}(Y_{ijk}|\, a_i, b_{ij}) $, showing that adding the
random intercept at the catch level is a way to preclude residual
over-dispersion of the model with only two random intercepts at the
village and site levels.
All environmental variables were first introduced in the model,
and a backward procedure was applied to select only those that
remained significant in the final model.
To achieve the most parsimonious model, adjacent classes of a
categorical variable were grouped together if the corresponding
regression estimates were close.\\

\textbf{ Predictive model}. A regression model was then constructed
to predict Anopheles count when only environmental data are
available.
The model was selected using a leave-one-out method (e.g. see
[26]). For a given set of covariates X, the following steps were
repeated for all catch sites j from 1 to 41:
\begin{enumerate}
 \item [i] A regression model of the number of Anopheles collected versus
environmental variables was performed using the observa-
tions from all sites except the i th (i.e. by excluding the 17 data
collections of the i th site)
\item [ii]This model was used to predict a number of Anopheles
collected P jk (k in 1... 17) in the 17 data collections at the i th
site using the corresponding known environmental covariates
for the i th site and the coefficients of the model from the
above step
\item [iii]The prediction errors $E_{jk}=|Y_{jk} - P_{jk}|/(Y_{jk}+1)$ were computed 
\end{enumerate}
Once the algorithm had been applied for the 41 sites, the median
of the 612 prediction errors (all catches at all sites) was determined,
and the final set of covariates was the one with the lowest median
prediction error.
After the selection of variables, interaction terms were
introduced and conserved in the final model if they led to a lower
median prediction error.
In this prediction model, correlation between the observations
was taken into account by entering a ‘‘village’’ variable in the
model. The general equation of this model was then:

\begin{equation}
 \label{equation_model_GLMM_pred}
  \ln [\mathbb{E}(Y_{ijk}| \beta)] = \beta_{0}+\sum_{l=1}^{p}\beta_{l}X_{ijkl}
\end{equation}

where $Y$  is the number of collected Anopheles, and $X$ is a vector of
environmental variables (including the ‘‘village’’ variable).
In order to evaluate our model’s ability to estimate the
spatiotemporal pattern of malaria transmission, we compared
the model’s predictions to the observed number of Anopheles
collected in the field.
We also used another predictive ‘‘pragmatic’’ model in which
the predicted number of collected Anopheles for the k th catch in the
j th site in the i th village was estimated by the mean number of
Anopheles collected at the other three sites in the same village during
the same catch, e.g. during the first survey in Gbetaga, 4, 7, 7 and 26 Anopheles were caught in the four catch sites respectively. The
numbers of collected Anopheles predicted by this pragmatic model
were then (7+7+26)/3, (4+7+26)/3, (4+7+26)/3 and (4+7+7)/3 in
the four houses respectively. Then, we compared the distribution
errors obtained from the two models (predictive and pragmatic)
according to the predicted Anopheles count.
All statistical analyses were performed by using STATA
software version 10 (Stata Corporation, College Station, Texas,
United States).
\section*{ Results}
During the 19 surveys between July 2007 and July 2009, a total
of 3,074 malaria vectors were caught (93.3
6.7
684 collections (19 catches at 4 catch sites in 9 villages), was 1
(interquartile range [0–4], max = 87). Evolution in vector density
as defined by the mean number of bites per human per night (m.a)
according to time is shown in Figure 1. These findings point to
time- and space-dependent fluctuations in vector density. Varia-
tion in m.a. was dependent on season and positively associated
with rainfall. Spatial differences in m.a. were observed between the
9 villages, particularly during each rainy season (from June to
November) even at a village scale, e.g. there was a strong
difference in m.a. changes between the 2 villages of Houngo and
Dohinoko which are only two kilometers apart (highlighted
curves), the first showing a low vector density throughout the
study and the second one showing strong seasonal variation with a
substantial increase during the rainy season. Furthermore, in all
villages (except Houngo) we observed marked m.a. differences
between catch sites, reflecting spatial variations in vector density at
the site level (Figure 2). As mentioned above, statistical analyses were conducted for just
17 surveys in which a total of 2,292 malaria vectors were collected.
Table 1 shows the final multivariate explanatory model. This
model contained random intercepts at the village, site and catch
levels, each random intercept improving the likelihood of the data.
Both the mean rainfall between 2 surveys and the number of rainy
days in the 10 days before the survey correlated positively with
Anopheles density, as expected. Independently, season also corre-
lated with Anopheles density, with density higher during rains.
Several site characteristics correlated with higher vector density:
proximity to a watercourse, a dry soil, and a higher NDVI
(vegetation index). All these results point to local spatiotemporal
variations in malaria transmission. Figure 3 shows a very good
adjustment between the number of Anopheles collected at each
survey and the explanatory model’s predictions.
Therefore, when entomological data are not available, a
predictive model based on environmental data could be useful to
estimate the spatiotemporal entomological risk in a house.
The best predictive model contained the following covariates:
season, mean rainfall between 2 surveys, number of rainy days in
the 10 days before the survey, use of insect repellent, NDVI, and
an interaction term between season and NDVI. Figure 4 shows
comparisons between the predictions generated with the regres-
sion model and the real number of Anopheles collected at each of
the 41 sites. The model fits with the actual spatiotemporal
transmission pattern for most but not all sites. Figure 5 shows a
comparison between the error distributions of the regression and
the pragmatic models, according to the real number of Anopheles collected. The error distributions and hence the predictive powers
of both models are highly comparable.
The number of infected Anopheles was low throughout the study:
the average Entomological Inoculation Rate (EIR) was 0.046 \begin{figure}[!h]
 \begin{center}
 \includegraphics[width=5.0in]{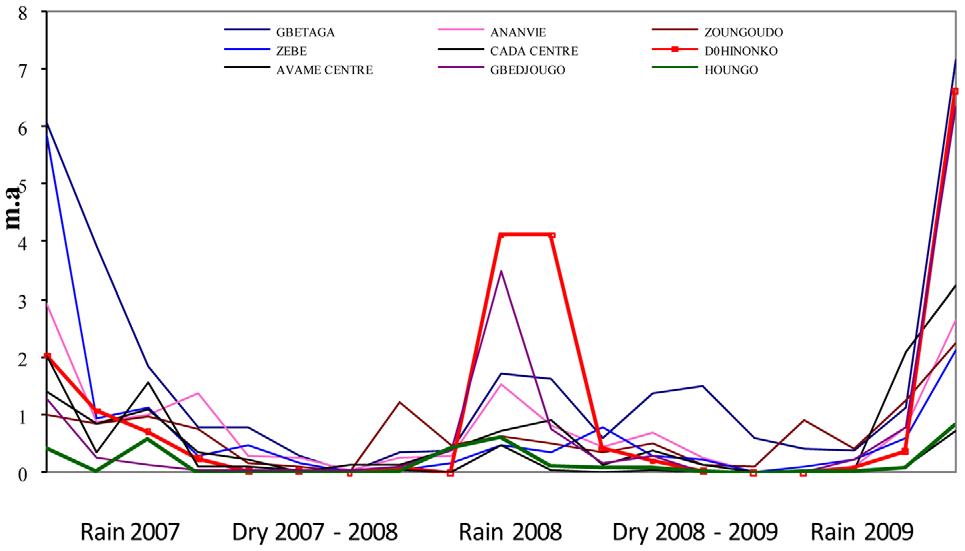}
\end{center}

\caption{\label{distribution_spatio_temporelle_village}\textbf{Number of Anopheles gambiae s.l. collected per man per day (ma) in the 9 villages for each of the 19 surveys.}}
\end{figure}
infected bite/human/night. When EIR was used as dependent
variable instead of m.a., the model failed to converge when too
many covariates were introduced together. However EIR and m.a.
were highly correlated (see figure 6, r = 0.95). Moreover, when
EIR was used as dependent variable with climatic variables—
mean rainfall between two collections, number of rainy days
during the 10 days before collection, and season—as the only
independent covariates, the same pattern was obtained (data not
shown). For these reasons we used the total number of Anopheles
caught on humans (m.a.) for the statistical analyses.

Discussion
This study set out to investigate the relationship between the
distribution of malaria vectors in southern Benin and local
environmental and climatic factors (at the level of both village
and house), and to propose a predictive model for the spatio-
temporal risk of exposure to Anopheles mosquitoes.
We observed substantial variations in malaria vector density at
the level of both village and house, even between houses which are
close together (separated by just a few dozen meters). We found  
that this variability could be explained not only by conventional
climatic factors (rainfall, season) but also certain environmental
factors, i.e. a watercourse nearby, and vegetation index and soil
type in the immediate surroundings (see also [27]).
The density of malaria vectors and the intensity of transmission
are relatively low in this area, confirming previous findings [8,22].
We showed that EIR and m.a. correlate strongly and, when EIR is
used as the dependent variable in our models, the pattern of the
results is the same. However, problems of stability and
convergence were observed with EIR so m.a. was used for the
statistical modeling. Nevertheless, there is no reason why an
infected mosquito’s behavior—which depends on ecological and
environmental conditions—would be any different [28], [29].
Based on this finding, we believe that our model based on Anopheles
density can accurately predict malaria transmission.
For statistical analyses, some continuous variables (e.g. NVDI,
rainfall levels...) were recoded as categorical variables, leading to a
loss of information. This loss of information is reduced by using
more classes and, furthermore, this method presents a double
advantage: the results are easy to interpret between the different
classes; and there is no need to assume a linear relationship

\begin{figure}[!h]
 \begin{center}
 \includegraphics[width=5.0in,height=5cm]{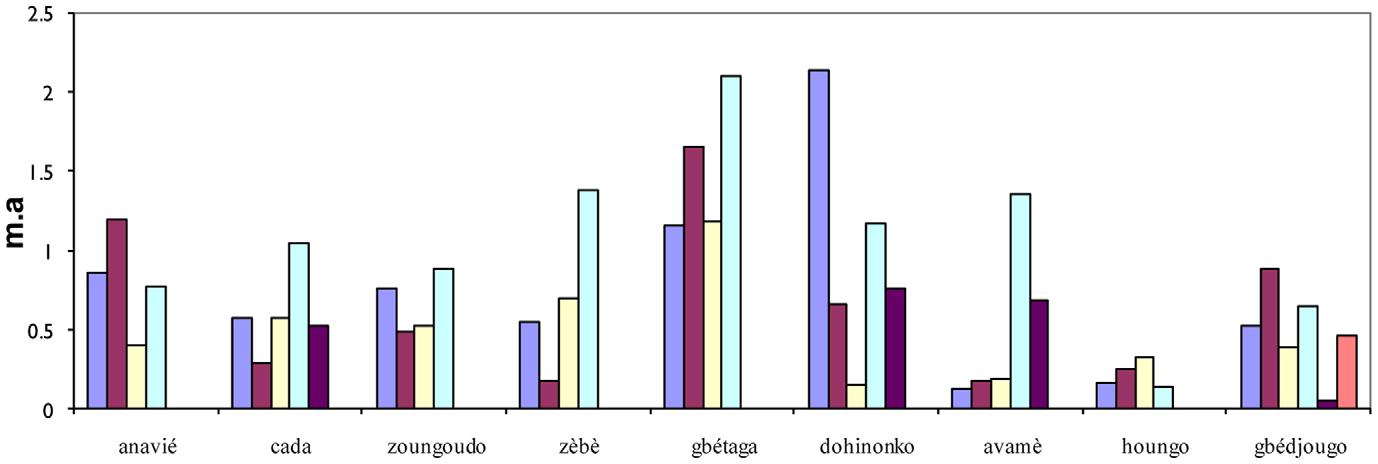}
\end{center}

\caption{\label{distribution_ma_village} \textbf{Mean m.a. in the 9 villages.} Each bar represents the mean m.a. throughout study in one house.}.
\end{figure}

\begin{table}[!h]
\caption{\textbf{Environmental factors associated with the density of malaria vectors at Tori Bossito, Benin (explanatory Poisson mixed
model)}} 
\begin{center}
\begin{tabular}{|p{0.2\textwidth}| p{0.25\textwidth} | p {0.15\textwidth} | p{0.15\textwidth} | p{0.12\textwidth} |} 
\hline
\textbf{Fixed effects}&\textbf{Classes} & \centering  Estimation&\centering Standard error &p-\textbf{value}\cr
\hline 
\hline 
Water-  & Yes&-&-&\cr
\cline{2-5}
course&No &1.869&0.63&0.003\cr
\hline 
Soil& Humid&-&-& \cr
\cline{2-5}
&Dry &2.27&0.72&0.002\cr
\hline
NDVI& Low&-&-& \cr
\cline{2-5}
&High &0.46&0.23&0.05\cr

\hline
Season& End of  dry season&-&-& $10^{-3}$\cr
\cline{2-5}
&Beginning of rainy season &1.63&0.18& \cr
\cline{2-5}
&End of rainy season &0.44&0.17& \cr
\cline{2-5}
&Beginning of dry season &60.49&0.19&\cr
\hline

Mean rainfall&Low&-&-& $10^{-3}$\cr
\cline{2-5}
&High&0.99&0.23& \cr

\hline
Number of raining days  & [0,1]&-&-&$10^{-3}$ \cr
\cline{2-5}
before collection &[2,4]&0.34&0.17& \cr
\cline{2-5}
& $>$ 4 &0.70&0.20& \cr
\hline
\end{tabular} \vspace{0.2cm}\\
\label{Tableau parametres fixes} 
\end{center}
\end{table}

\begin{table}[!h]
\caption{\textbf{Random effects estimation}} 
\begin{center}
\begin{tabular}{|p{0.2\textwidth}| p{0.25\textwidth} | p {0.2\textwidth} | p{0.21\textwidth} |} 
\hline
\textbf{Random effects}&\textbf{Level} &\textbf{Estimation} &\textbf{Standard error}\cr
\hline 
\hline
Random  & village&0.71&0.19 \cr
\cline{2-4}
intercept& House&0.21&0.11 \cr
\cline{2-4}
& caught&1.04&0.06 \cr
\hline
\end{tabular} \vspace{0.2cm}\\
\label{Tableau parametres fixes} 
\end{center}
\end{table}

between the dependent variable and the covariate. Three random
intercepts were introduced in our explanatory model, at the
village, site, and catch levels. Each random intercept increased
significantly the likelihood of the data (with the highest increase
afforded at the catch level). This is consistent with the fact that
they correctly take into account the hierarchical structure of the
data, as well as unobserved variables which could explain the
vectors variability at each of the three levels. The quality of the  models adjustment to the data was spectacularly improved by the
random intercept at the catch level. This makes it possible to
introduce this intercept to deal with over-dispersed data—as
recommended by Rabe-Heskett and al. [30]—and confirms the
reliability of the model.
The association between vector density and environmental or
climatic factors has been widely studied [12,13,31,32,33,34] with
rainfall and season consistently identified as significant factors. We  \begin{figure}[!h]
 
 \begin{center}
 \includegraphics[width=5.0in]{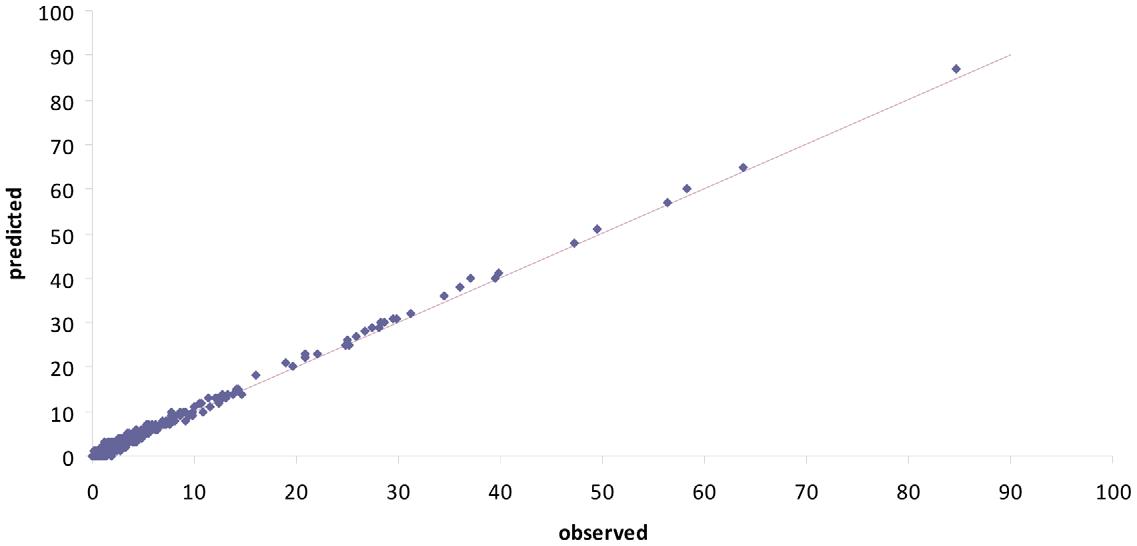}
\end{center}
\caption{\textbf{Relationship between observed and predicted numbers of Anopheles gambiae collected (explanatory model).} The straight
line is the bisector. \label{Comparaison_obs_pred_mod_explicatif}}
\end{figure}

 \begin{figure}[!h]
 \begin{center}
 \includegraphics[width=5.0in]{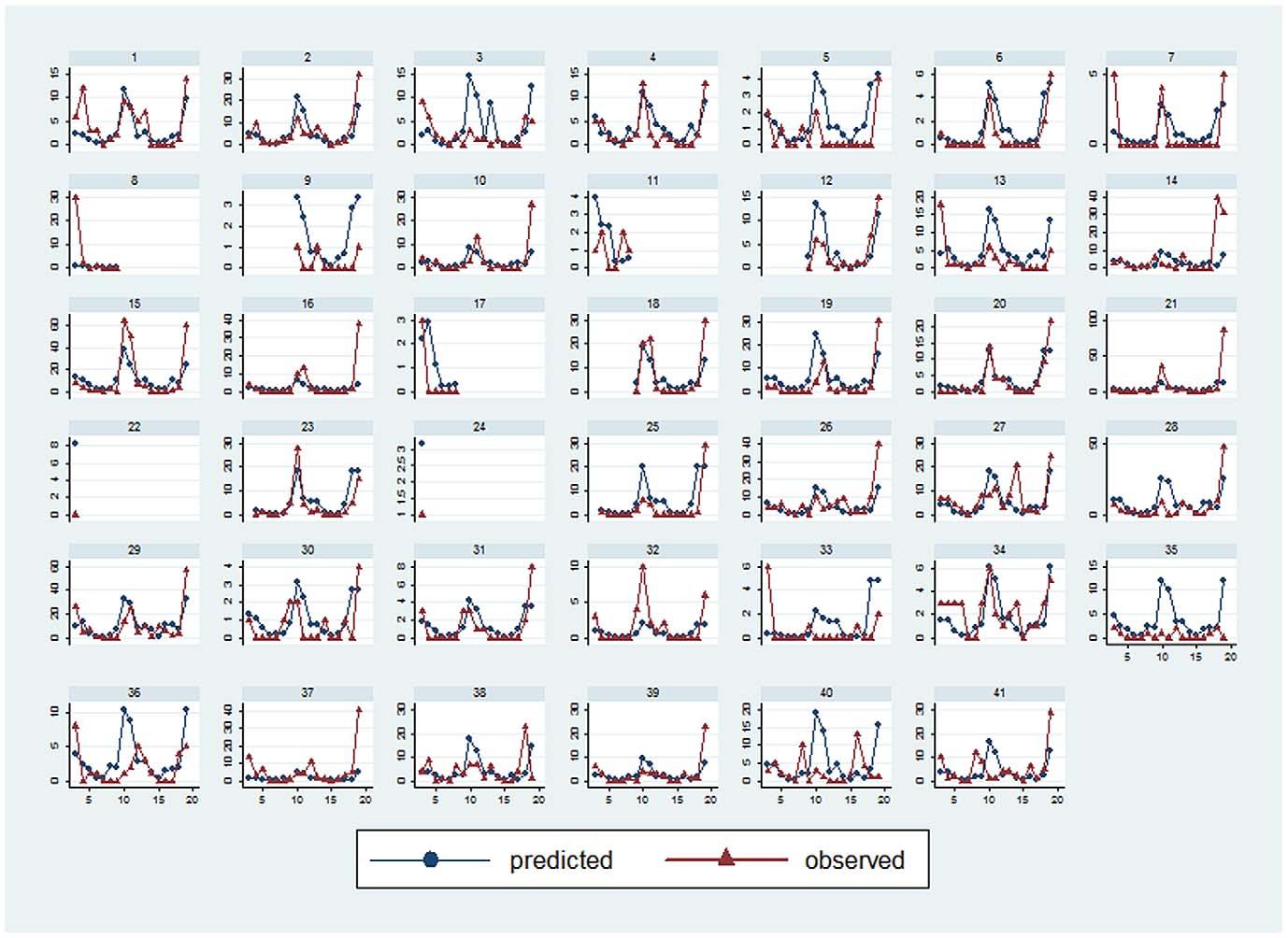}
\end{center}

\caption{\textbf{(Predictive model) and observed numbers of Anopheles in the 41 houses. } Each graph shows the observed (solid line) and the
predicted (dashed line) number of Anopheles during each catch in a house. \label{comparaison_pred_obser_41_maisons}}
\end{figure}

have identified the number of rainy days before mosquito collection
as an additional factor: this could be explained by an increase in the
number of temporary water habitats (puddles) favorable to the
development of the mosquito larvae. There was no correlation
between m.a. and the presence of a net or repellent use in the house,
even when the indoor biting rate was considered. This could be
explained by the fact that we used a man-landing capture technique
without a bed net in a limited number of houses. We also showed
substantial variations in vector density, not only between different
villages but also in the same village, between houses within a few
dozen meters of one another. Factors that could explain this include
factors in the houses’ immediate surroundings: the presence of a
watercourse nearby, a higher vegetation index, and dry soil were all
associated with higher vector density. The positive correlation with
dry rather than hydromorphic soil could be explained by the fact
that the latter is concentrated around the main river in the area, and
is overrun by dense aquatic vegetation which prevents Anopheles
gambiae breeding.
Other studies that have investigated the relationship between
domestic features and malaria transmission [35,36] have shown
that roof and ceiling type can also affect malaria transmission. All
these observations point up the importance of taking local
characteristics—of the village, the house and the house’s
immediate surroundings—into account when dealing with the
variability of malaria transmission. These findings may also have important consequences when
focusing on mechanisms to explain differences in P. falciparum
infection or the incidence of malaria attacks between groups of
individuals when both ‘‘environmental’’ and ‘‘biological’’ deter-
minants are involved. This is particularly relevant for children
born to a mother with PAM who tend to develop malaria sooner
[16,17,18,19]. The hypothesis of immune tolerance has been put
forward to explain this, i.e. fetal exposure to malaria modulates
neonatal immunity in endemic areas where infection during
pregnancy is common [37,38]. However, even though immune
tolerance is due to infection status of the mother during gestation,
its consequences in term of newborn’s susceptibility to infection
manifests itself through the fact that offspring of mothers with
placental malaria at delivery experience their first P. falciparum
parasitemia at a younger age. However, it also seems self-evident
that babies more strongly exposed to Plasmodium would be at
greater risk of contracting malaria rapidly after birth. In
consequence, to demonstrate immune tolerance and evaluate its
role in determining susceptibility to early malaria infection, local
transmission variations have to be taken into account. This
parameter can be addressed by a statistical method that takes into
account both spatial and temporal variability but, in the studies
cited above, the only exposure-related variable introduced into the
Cox model was ‘‘area of residence’’. Such a variable—which is
time-independent and has the same value for all children living in

 \begin{figure}[!h]
 \begin{center}
 \includegraphics[width=5.0in]{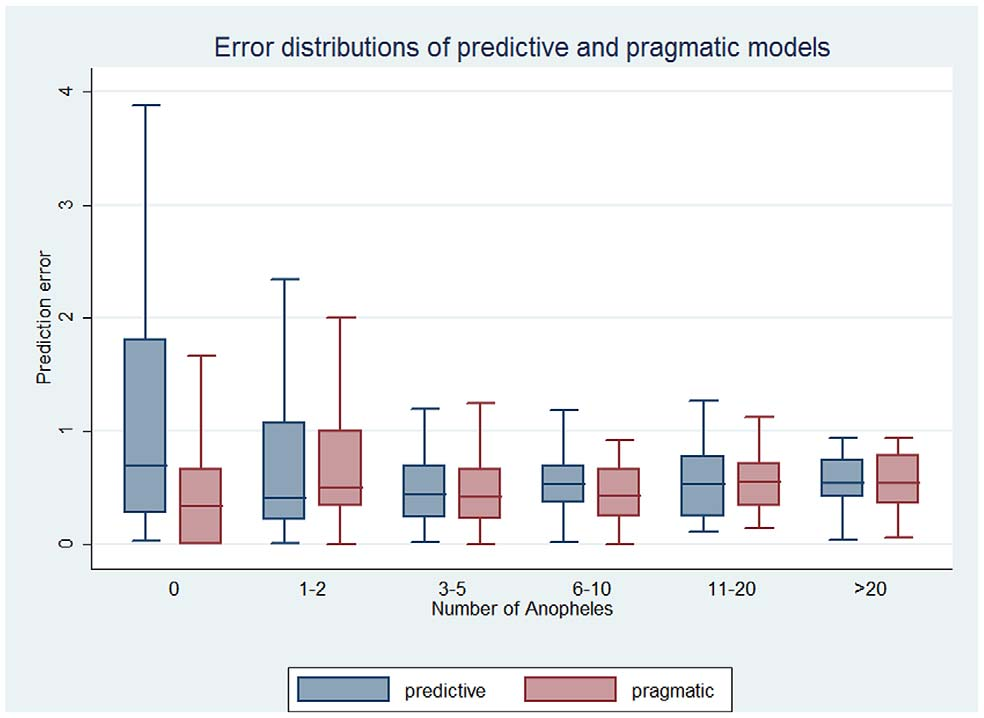}
\end{center}
\caption{ \textbf{Error distributions of the pragmatic and predictive models according to the number of observed Anopheles.}   In each group
(number of Anopheles), the left box corresponds to the predictive regression model and the right box to the pragmatic regression model.\\
\label{distribution_erreur_pred_expl_prag}} 
\end{figure}
the same village—provides little information about the differential
exposure of children in a cohort. After initial demonstration of
such local spatiotemporal variability, a novel approach was
formulated to predict malaria transmission in houses within a
limited area with known ecological and environmental character-
istics. We have demonstrated that a regression model including
spatial and time-dependent variables at the village and the house
levels, yields a spatiotemporal prediction of malaria transmission
comparable to that obtained on the basis of entomological data.  This approach constitutes a substantial improvement and the
model will be applied to all children in the cohort over the relevant
period. The predictions will be used as a time-dependent covariate
in analysis of the interval before the first malaria infection (Cox
model) to investigate the role of immune tolerance in this
parameter.
Finally, this approach can be used to estimate spatiotem-
poral variations in malaria transmission in cohort studies,
thereby delineating and elucidating the respective roles of
\begin{figure}[!h]
 \begin{center}
  \includegraphics[width=5.0in]{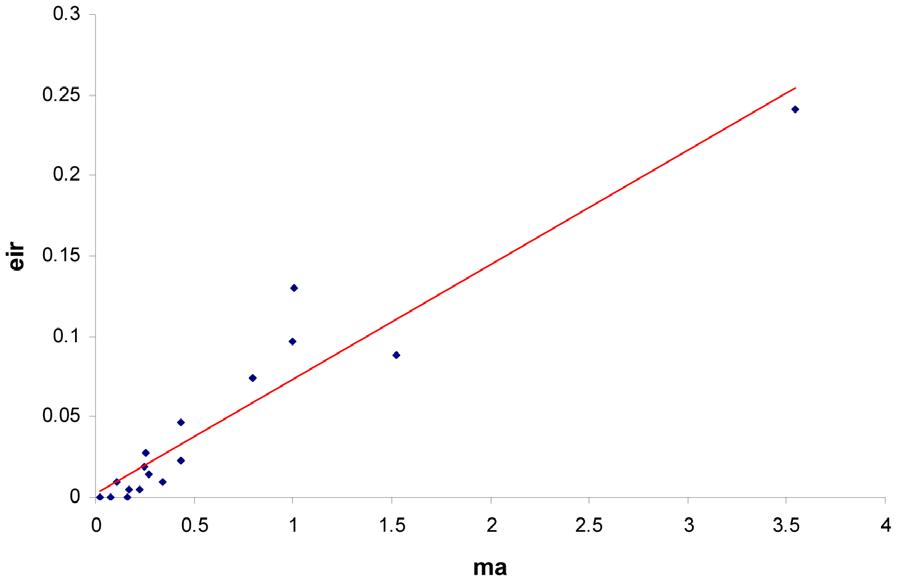}
\end{center}
\caption{ \textbf{Relation entre la moyenne des m.a. et l'EIR} \label{distribution_erreur_pred}}. Sur l'axe des abscisses la moyenne des m.a. 
pour toutes les maisons durant une mission et en ordonn\'ee la moyenne des EIR pour 
toutes les maisons durant la m\^eme mission.
\end{figure} 

environment, behavior and physiology in determining suscep-
tibility to infection.

\section*{ Acknowledgments}
We thank the entire population of Tori Bossito, the staff of mosquito
collectors and the laboratory team.

\section*{Author Contributions}
Conceived and designed the experiments: GC CP AG. Performed
experiments: GC BK CP AB AG. Analyzed the data: GC BK MNH
AG. Contributed reagents/materials/analysis tools: GC BK CP AlP
MNH NF AM VC AG. Wrote the paper: GC BK CP AlP AB MNH
AM VC AG.

\section*{References}

1. Barnes KI, Chanda P, Ab Barnabas G (2009) Impact of the large-scale
deployment of artemether/lumefantrine on the malaria disease burden in Africa:
case studies of South Africa, Zambia and Ethiopia. Malar J 8 Suppl 1: S8.\\
2. Ceesay SJ, Casals-Pascual C, Erskine J, Anya SE, Duah NO, et al. (2008)
Changes in malaria indices between 1999 and 2007 in The Gambia: a
retrospective analysis. Lancet 372: 1545–1554.\\
3. O’Meara WP, Bejon P, Mwangi TW, Okiro EA, Peshu N, et al. (2008) Effect of
a fall in malaria transmission on morbidity and mortality in Kilifi, Kenya. Lancet
372: 1555–1562.\\
4. WHO (2009) World Health Organisation - World Malaria Report.\\
5. Guerra CA, Gikandi PW, Tatem AJ, Noor AM, Smith DL, et al. (2008) The
limits and intensity of Plasmodium falciparum transmission: implications for malaria
control and elimination worldwide. PLoS Med 5: e38.\\
6. Guthmann JP, Llanos-Cuentas A, Palacios A, Hall AJ (2002) Environmental
factors as determinants of malaria risk. A descriptive study on the northern coast
of Peru. Trop Med Int Health 7: 518–525.\\
7. Ageep TB, Cox J, Hassan MM, Knols BG, Benedict MQ, et al. (2009) Spatial
and temporal distribution of the malaria mosquito Anopheles arabiensis in northern
Sudan: influence of environmental factors and implications for vector control.
Malar J 8: 123.\\
8. Djenontin A, Bio-Bangana S, Moiroux N, Henry MC, Bousari O, et al. (2010)
Culicidae diversity, malaria transmission and insecticide resistance alleles in
malaria vectors in Ouidah-Kpomasse-Tori district from Benin (West Africa): A
pre-intervention study. Parasit Vectors 3: 83.\\
9. Smith DL, Dushoff J, Snow RW, Hay SI (2005) The entomological inoculation
rate and Plasmodium falciparum infection in African children. Nature 438:
492–495.\\
10. Smith DL, McKenzie FE, Snow RW, Hay SI (2007) Revisiting the basic
reproductive number for malaria and its implications for malaria control. PLoS
Biol 5: e42.\\
11. Killeen GF, Smith TA (2007) Exploring the contributions of bed nets, cattle,
insecticides and excitorepellency to malaria control: a deterministic model of
mosquito host-seeking behaviour and mortality. Trans R Soc Trop Med Hyg
101: 867–880.\\
12. Dery DB, Brown C, Asante KP, Adams M, Dosoo D, et al. (2010) Patterns and
seasonality of malaria transmission in the forest-savannah transitional zones of
Ghana. Malar J 9: 314.\\
13. Mourou JR, Coffinet T, Jarjaval F, Pradines B, Amalvict R, et al. (2010) Malaria
transmission and insecticide resistance of Anopheles gambiae in Libreville and Port-
Gentil, Gabon. Malar J 9: 321.\\
14. Bojang KA, Olodude F, Pinder M, Ofori-Anyinam O, Vigneron L, et al. (2005)
Safety and immunogenicty of RTS,S/AS02A candidate malaria vaccine in
Gambian children. Vaccine 23: 4148–4157.\\
15. Dicko A, Diemert DJ, Sagara I, Sogoba M, Niambele MB, et al. (2007) Impact
of a Plasmodium falciparum AMA1 vaccine on antibody responses in adult
Malians. PLoS One 2: e1045.\\
16. Le Hesran JY, Cot M, Personne P, Fievet N, Dubois B, et al. (1997) Maternal
placental infection with Plasmodium falciparum and malaria morbidity during the
first 2 years of life. Am J Epidemiol 146: 826–831.\\
17. Mutabingwa TK, Bolla MC, Li JL, Domingo GJ, Li X, et al. (2005) Maternal
malaria and gravidity interact to modify infant susceptibility to malaria. PLoS
Med 2: e407.\\
18. Schwarz NG, Adegnika AA, Breitling LP, Gabor J, Agnandji ST, et al. (2008)
Placental malaria increases malaria risk in the first 30 months of life. Clin Infect
Dis 47: 1017–1025.\\
19. Bardaji A, Siguaque B, Sanz S, Maixenchs M, Ordi J, et al. (2011) Impact of
Malaria at the End of Pregnancy on Infant Mortality and Morbidity. The
Journal of Infectious Diseases. pp 691–699.\\
20. Ellman R, Maxwell C, Finch R, Shayo D (1998) Malaria and anaemia at
different altitudes in the Muheza district of Tanzania: childhood morbidity in
relation to level of exposure to infection. Ann Trop Med Parasitol 92: 741–753.\\
21. Sylla EH, Lell B, Kun JF, Kremsner PG (2001) Plasmodium falciparum
transmission intensity and infection rates in children in Gabon. Parasitol Res
87: 530–533.\\
22. Damien GB, Djenontin A, Rogier C, Corbel V, Bangana SB, et al. (2010)
Malaria infection and disease in an area with pyrethroid-resistant vectors in
southern Benin. Malar J 9: 380.\\
23. Gillies M, De Meillon B (1968) The Anophelinae of Africa south of the Sahara).
Pub South Afr Inst Med Res Johannesburg.\\
24. Gillies M, De Meillon B (1987) A supplement to the Anophelinae of Africa south
of the Sahara (Afrotropical region). Pub South Afr Inst Med Res.\\
25. Wirtz RA, Zavala F, Charoenvit Y, Campbell GH, Burkot TR, et al. (1987)
Comparative testing of monoclonal antibodies against Plasmodium falciparum
sporozoites for ELISA development. Bull World Health Organ 65: 39–45.\\
26. Efron B, Tibshirani R (1993) An introduction to the bootstrap; Chapman $\&$
Hall/CRC, editor . 436 p.\\
27. Pierrat C (2010) Pour un diagnostic territorial d’un risque sanitaire \`a l'\'echelle fine:
essai de mod\'elisation des variables environnementales du paludisme (Sud du
B\'enin. Revue Espaces Tropicaux; In Press.\\
28. Lacroix R, Mukabana WR, Gouagna LC, Koella JC (2005) Malaria infection
increases attractiveness of humans to mosquitoes. PLoS Biol 3: e298.\\
29. Hurd H (2010) Parasite-mediated enhancement of transmission by hematopha-
geous insects. Wageningen: W. TAKKEN and B. KNOLS Wageningen
Academic Publisher. pp 349–364.\\
30. Rabe-Hesketh S, Skrondal A (2008) Multilevel and Longitudinal Modeling
Using Stata, 2nd Ed Stata Press. 562 p.\\
31. Briet OJ, Vounatsou P, Gunawardena DM, Galappaththy GN, Amerasinghe PH
(2008) Models for short term malaria prediction in Sri Lanka. Malar J 7: 76.
32. Haghdoost AA, Alexander N, Cox J (2008) Modelling of malaria temporal
variations in Iran. Trop Med Int Health 13: 1501–1508.\\
33. Olson SH, Gangnon R, Elguero E, Durieux L, Guegan JF, et al. (2009) Links
between climate, malaria, and wetlands in the Amazon Basin. Emerg Infect Dis
15: 659–662.\\
34. Bui HM, Clements AC, Nguyen QT, Nguyen MH, Le XH, et al. (Epub 2010
Sep 15) Social and environmental determinants of malaria in space and time in
Viet Nam. Int J Parasitol 41: 109–116.\\
35. Ye Y, Hoshen M, Louis V, Seraphin S, Traore I, et al. (2006) Housing
conditions and Plasmodium falciparum infection: protective effect of iron-sheet
roofed houses. Malar J 5: 8.\\
36. Atieli H, Menya D, Githeko A, Scott T (2009) House design modifications
reduce indoor resting malaria vector densities in rice irrigation scheme area in
western Kenya. Malar J 8: 108.\\
37. Cot M, Le Hesran JY, Staalsoe T, Fievet N, Hviid L, et al. (2003) Maternally
transmitted antibodies to pregnancy-associated variant antigens on the surface of
erythrocytes infected with Plasmodium falciparum: relation to child susceptibility to
malaria. Am J Epidemiol 157: 203–209.\\
38. Malhotra I, Dent A, Mungai P, Wamachi A, Ouma JH, et al. (2009) Can
prenatal malaria exposure produce an immune tolerant phenotype? A
prospective birth cohort study in Kenya. PLoS Med 6: e1000116.

\end{document}